Third time is the charm – Why the World just might be ready for the Internet of Things this time around.

Peter M. Corcoran
College of Engineeing & Informatics
National University of Ireland Galway
Galway, Ireland
peter.corcoran@nuigalway.ie

*Abstract*— The technology to connect 'things' to the Internet has existed for more than 20 years, so if we take a look back at recent history we might well be tempted to ask the question why will IoT 'happen' this time around. In this paper we examine the origins of the Internet of Things, answer the question "Why Now?", and look forward to the next wave of disruptive technologies that will be coming to a device near you in the next few years.

*Index Terms*—Internet-of-Things, IoT History.

## I. A Short History of the Internet

The term "Internet of Things" (IoT) was first documented by British visionary, Kevin Ashton, in 1999. He used the phrase to describe a system where the Internet connects to the 'real world' via an ubiquitous network of data sensors. Of course the use of this term has grown somewhat beyond the original intention and today it means many things to many people. But to get back to the root of it all we should also consider the "Internet" itself in order to understand the full context of the IoT.

The origins of the Internet go back to the Arpanet in the late 1960's. By 1970 there were five permanent nodes on the Arpanet at several of the largest US. Learning from these early days the researchers realized they needed to build a robust data protocol that could recover from transmission errors. In 1974 Vint Cerf introduced TCP/IP but it was a decade later before it was broadly adopted across the network and the real growth could start. In 1984 the c.1,000 active network nodes on the early "Internet" switched over to adopt TCP/IP for their core data transmission and networking protocol. And since then the network has continued to grow unabated.

There are two key things to remember about the Internet: (i) the "Internet" is not the Web; you can think of the Web as a GUI for the display & publishing of data carried by the Internet but the underlying data transports, in particular TCP/IP, are what have allowed the Internet to scale; (ii) the "Internet" was designed to military specifications as a 'battlefield' protocol; it is designed to be able to adapt to unreliable channels and to recover from data loss.

This last point has allowed the "Internet" to grow consistently over the last 4-5 decades and the introduction of mobile devices has further driven this demand. In fact you could say that today a computer is pretty useless without Internet connectivity and that underlying connectivity is provided by TCP/IP.

## II. A Personal Perspective on the early IoT

The technology to connect 'Things' to the Internet has actually existed for more than two decades. I can confirm this personally as I was blissfully connecting CEBus devices to the Internet 3-4 years before Ashton's revelations [1], [2] and even developing Java user-interfaces for these 'Things' [3]. But let us begin this journey back in time with a workshop I gave in 2002 at the IEEE International Conference on Consumer Electronics (ICCE 2002) entitled:.

A. ICCE 2002 – Home Networks for the 21$^{st}$ Century

This tutorial session that I organized covered a range of topics from the physical layers, to connectivity technologies available at that time and implementing TCP/IP on 8-bit embedded microcontrollers. The conclusion at that time was that TCP/IP was the way to glue things together at the lower layers of the communications stack but there was a need for a *middleware* layer to sit on top that would understand the nature and capabilities of individual devices. The final presentation in the tutorial considered OSGi and UPnP for this middleware, concluding that UPnP was a good foundation, but needed additional capabilities, especially for the UI.

In fig 1 you'll note a slide from this tutorial giving an overview on how to build a 'Thing'; note that at that time 16-bit micros were state-of-art and low cost devices would have likely employed 8-bit micros. And the optimal connectivity technologies for low-cost devices were Bluetooth for wireless, which was difficult to use, or wired Ethernet.

- HOWTO build a low-cost home network peripheral:
  - Network Connection Methods (incl. costs)
    - Ethernet, Bluetooth, Powerline, Dial-Up, WLAN
  - Useful 8/16-bit Microcontrollers
    - PIC, Ubicom (Scenix), …
  - Practical Solutions:
    - Bluetooth/Ethernet Bridge
    - Low cost Ethernet Solutions
- Description of latest Powerline Technology (14 Mbps)
  - How it works; technical overview; products now available!
- Overview of IEEE 1394/802.11b Internetworking
  - How it works; advantage & reasons for wireless bridging

Fig. 1. How-to build a 'Thing' in 2002

This was based on the cost factors shown in Figure 2. Note the cost of a hardware Ethernet was low enough to be considered for mass-market products and Bluetooth was almost cheap enough to be a contender. Wifi technology was far too expensive at this time for genuine consumer products, being 10's of USD even in high-volumes.

| Network Type | Speed | Wiring Needs | Production Cost | End-User Cost |
|---|---|---|---|---|
| Bluetooth | < 1 Mbps | None | <$8 | $75-$100 |
| Ethernet | 10/100 Mbps | Category 5 UTP | $2-$15 | $25-$50 |
| Phone-line | 10 Mbps | existing phone | $20-$30 | $45-$75 |
| Power-line (Intelogis) | 50-350 Kbps | existing electrical | $15 | $25 |
| Power-line (IntelIon) | 14 Mbps | existing electrical | $25 | $100 |
| Wireless Ethernet (SWAP) | 1-2 Mbps | None | $25 | $70-$200 |
| Wireless Ethernet (Wi-Fi) | 11 Mbps | None | $30-$45 | $100-$300 |

Fig. 2. The costs of connectivity in 2002.

At this point you might start to wonder why the IoT didn't 'happen' in 2002. The enabling technologies were clearly available and there were many people interested in how to developed connected devices – we had strong attendance at this tutorial. But somehow the pieces didn't fit together in a compelling way. Over the following decade it became clear to me that simply connecting "Things" to the Internet simply doesn't create enough value to sustain practical business models – if it did then IoT would have happened back in 2002.

B. The Age of CEBus – the mid-1990's

If you know what CEBus [4]–[6] is then you are likely a CE industry veteran like myself. It was a networking standard developed originally by the Electronic Industries Association (EIA) and then adopted by the Consumer Electronics Association (CEA). The standard was very much ahead of its time and supported multiple physical layers including twisted-pair, coaxial cable, powerline, wireless and even RF.

I had discovered CEBus in the early 1990's and managed to find resources to purchase some development kits – it was my first exploration of advanced networking protocols on embedded controllers. At the time there was also a new OS based on Unix available called Linux and in my youthful enthusiasm I had started playing with this new toy. It was a breath of fresh air to be able to compile the OS from scratch. And of course it also gave access to the TCP/IP stack. Perhaps you can now begin to understand how I ended up delivering that tutorial nearly a decade later in 2002?

It seemed natural to build a Linux-based 'gateway' between our CEBus nodes and the Internet 1], [2]. Then a new technology called Java arrived on the scene – it was network-centric and provided a powerful UI capability, for that time. Again it made sense to build UI-components that would be activated by the presence of CEBus devices [3]. Not only could we connect devices to the Internet, but they could have dynamic user interface elements so people could control and interact with these 'Things'.

C. Waiting for GODOT

Apologies to Samuel Beckett, but I must confess that I've been waiting to "get our devices out there" (GODOT) for more than two decades. In the beginning I was so keen I even set up a small company in the late 1990's with a view to developing the next generation of connected consumer devices. That company eventually moved in a different direction (www.fotonation.com) but anecdotally, its original business was "connectivity for devices" which then transformed into "connectivity for cameras" and continued to grow and evolve towards the technology needs of the nascent digital consumer imaging industry.

But I digress a bit. It is clear that the enabling technology for IoT existed not only in 2002, but in fact 6-7 years earlier in the mid-1990's it was also available and arguably in a more capable, functional and scalable form than many of today's IoT solutions. So if the technology existed and there has not really been any disruptive breakthrough then we arrive at the same question - why did IoT not go mainstream before now?

III. IoT in 2016 – What is Different?

The world wasn't ready for IoT in the mid-1990's or in the early 2000's, so what has changed in the meantime? If you know a little about me you may already know that I have been an IoT skeptic in recent years, but now I begin to see some things that are changing my views. To get a better understanding and context lets recap what we know hasn't changed a lot:

*The Internet* is still pretty much the same; it got bigger and more technologies have been layered on top of TCP/IP and its little brother, UDP. We have a lot of support now for audio and video traffic which shows that the capacity and capability of the infrastructure has increased, but there have not been any radical changes – just a constant growth of nodes and data traffic.

*Embedded devices* haven't changed radically either. Yes, we have moved to 32 bit systems and most embedded devices can easily support a sophisticated OS, but the underlying connection is still via a TCP/IP stack which was available nearly 20 years ago for 8/16 bit devices.

*Connectivity technologies* have improved but not radically. Yes, we can now have more sophisticated Wifi connections for less than 5 USD, but you could have achieved a wireless link in 2002 using Bluetooth and an Ethernet bridge. There isn't anything very disruptive here that would argue that IoT will become a commercial success today.

So now lets consider what is new and has changed the technology landscape.

A. The Smartphone

We live in an age of persistent connectedness and are increasingly empowered as individuals to generate large amounts of digital data. Increasingly our smartphone is the go-to hub of our increasingly connected digital lives. Smartphones are used for a continuously expanding array of applications, from Internet browsing to e-mailing, to gaming, to banking, to shopping, and managing travel arrangements – airfares, car rental, etc. And for many of us they have become the primary tool to record and document our personal lives in pictures and video - a connected gateway that you carry with you all day long. The 'smarter' and more capable these devices become, the more they infiltrate our daily activities and blend themselves into our personal lives.

B. Cloud Computing

The second change in the technology landscape is the emergence of 'the Cloud'. All that data created by your smartphone has to go 'somewhere' in order to create 'value'. I've written elsewhere about the synergies between CE and 'the Cloud' [7]–[11]. Cloud computing is another concept that originated in the 1960's and has only became a reality in the last decade. For most of us 'the Cloud' has also been slowly infiltrating itself into our daily activities. Most of us have at least one 'cloud mail' account – I have 7 separate accounts on Gmail, each with its own purpose – a different face on my complex work & personal life(s). Many of us also make use of file sharing services such as Dropbox, iCloud, Google Drive. And did I mention photos & video? Individual consumers are now the primary generators of content for "the Cloud" [7]. And in that context, video and images are the main drivers of storage & infrastructure.

C. Mobile Devices & Data Networks

Smartphones have, in turn, created the demand for today's mobile networks. In in Figure 3 we show a table with the data transfers per month from a range of devices. The table is a few years old [12] but it captures the spirit of data growth on the Internet.

| Device Category | 2009 | 2010 | 2011 | 2012 | CAGR (2012-2017) | Projected (2017) |
|---|---|---|---|---|---|---|
| Laptop | 1,145 | 1,460 | 2131 | 2,503 | 31% | 5,731 |
| Smartphone | 35 | 55 | 150 | 342 | 81% | 2,660 |
| Smartphone (4G) | -- | -- | -- | 1,302 | -- | 5,114 |
| Tablet | 28 | 405 | 517 | 820 | 113% | 5,387 |
| Gaming Console | -- | 244 | 317 | -- | -- | NA |
| Mobile Phone | 1.5 | 1.9 | 4.3 | 6.8 | -- | 31 |

Data use in MB per month; note the CAGR rates of 81% and 113% for smartphones & tablets; Tablets will consumer as much "Network Data" as Laptops by 2017 – 2x the consumption of today's laptops; but there will be a lot more tablets & smartphones …

Thin Clients like Smartphones will drive Data Consumption … and Production! (via Pictures, Videos, etc) …

Fig. 3. Growth in data traffic generated by mobile devices.

If you look at the 2012 numbers in Figure 3 you see that a laptop was generating about 2.5 GB of network data; by 2017 it was predicted to be generating nearly 6 GB. But the real story lies with the 4G smartphones and tablets both of which will be over the 5 GB threshold by 2017. Now the real story is that there were only 0.6 billion laptops in 2012 and market growth is low. In contrast there will be of the order of 2B active tablets by 2017 and likely 4-5 billion smartphones.

So to sum up these points:
- The Cloud has evolved a set of *sophisticated infrastructures* for storage, messaging, security, content & connectivity
- Mobile networks have driven *ubiquitous connectivity*
- Smartphones provide the *user interface* (and a *gateway* for some devices) to access, manage and control our "Things"
- And the Internet means this new infrastructure is accessible everywhere, genuinely ubiquitous.

IV. What Comes Next?

You don't have to go far to find some pretty amazing predictions for the adoption of 'things'; Gartner has predicted 26 billion units by 2020; Cisco has an even higher estimate of 50 billion. Of course it depends on (i) what you consider to be a "Thing" and (ii) what you consider to be 'connected to the Internet'. If we include devices connected to secondary networks such as Bluetooth, RFID nodes and Home Networks such as Zigbee, 6LoPan and others, well then these estimates start to make some sense. Data can certainly make its way from such secondary networks onto the Internet.

So if IoT does happen – and it looks that everything is now in place for that to happen – then it is going to be BIG! Maybe even BIGGER than the smartphone revolution? You can find a lot more examples of different examples of 'things' in the last part of my IEEE webinar at: *http://j.mp/PC_IoTWebinar*

But there will also be some challenges. Lets take a quick look at some of these before we move to conclude that the time for GODOT has finally arrived.

V. Challenges and Scary Stories

A. Privacy & CyberSecurity

So when every device is connected and equipped with a wide range of sensing technologies how will be manage and preserve individual privacy? As cameras grow smaller and smaller and wearable technologies become practical how will you know who is recording your meetings and interactions with other persons?

As editor-in-chief of IEEE Consumer Electronics Magazine I've seen an increasing number of articles discussing these issues. Examples include the use of Google Glass to observe and learn user PIN numbers; the NEST thermostat which can be hacked and knows when you are, and aren't at home; home security and baby monitor cameras that are easily hacked and in many cases they stream open, un-encrypted video data in well-known H264/MPEG formats. Anyone with a moderate technical skill level can intercept you home security video and learn if you are home!

Most connected devices are secured with a factory-supplied default username + password. Users rarely change these as they will 'have to remember" a new username and password. So to hack many devices you simply log on as 'admin' and type in

'password'; on other devices these default values are written 'on the box'. So our first major concern is that of privacy and cyber-security – the two are intertwined and properly designed security protocols will support and benefit privacy on a device.

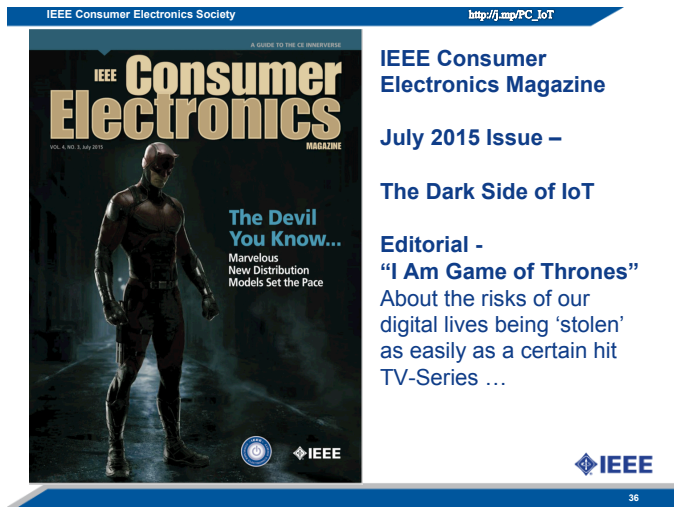

Fig. 4. IEEE CE Magazine has seen increasing growth in articles addressing a variety of privacy & cybersecurity topics.

B. Energy Issues

There are going to be many different types of 'things' but we can be sure of one aspect of each type of 'thing' – they will all use energy. The amount will vary, but the reality is that devices must either be connected to a fixed power line or incorporate a battery. The former implies a fixed installation and potentially a new wired power infrastructure will be needed to match the requirements of these 'things'; the latter implies another device that will need to have its battery charged or changed on a regular basis – definitely a new barrier to broad adoption of the IoT.

But the energy use of the 'things' themselves is only part of this equation – to accommodate a growing number of IoT devices the existing wireless and network infrastructures will have to expand, and as most of these devices will use wireless connectivity for convenience it is the wireless communications infrastructure that has to grow, and potentially grow quite rapidly. And it is this same infrastructure that is already the main driver of global electricity consumption for consumer ICT [15].

Figure 5 shows how the balance between operational energy usage, networks and data centers will change under the influence of smartphone & tablet growth up to 2017. Note how network energy increases from 20% to nearly 30% over a few short years. If the activity of 3-4 billion smartphones can cause such a shift then 50 billion 'things' is going to impact energy consumption of the network infrastructure by a similar or larger measure so we could be moving towards the era where 50+% of energy is due to the network! Remember that many 'things' will run 24/7 and consume energy continuously. Even if the devices themselves are quite low power they need a communications infrastructure that is not so low-power and in many cases a cloud data service that also can use significant amounts of energy.

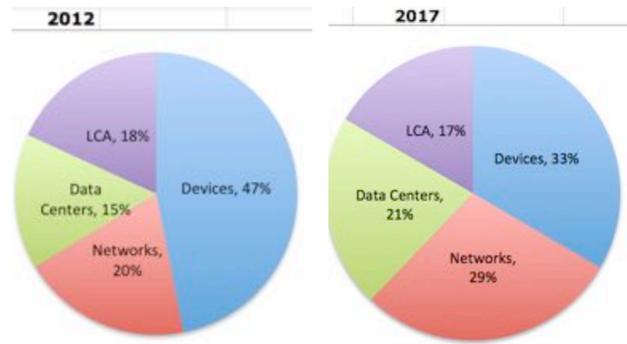

Fig. 5. Growth in data traffic generated by mobile devices.

VI. Concluding Thoughts

As a long-term IoT skeptic I recently agreed to give an IEEE Webinar on this topic (*http://j.mp/PC_IoTWebinar* ). As a consequence I had to examine and re-think many of my arguments based on the last 2 decades that I spent "waiting for the IoT". In this article I've explained how I came to a modified view on the Internet of Things. Yes, it may well be happening, driven by a combination of improved "Cloud" infrastructure, the smartphone revolution and recent improvements in mobile data networks. There are still multiple barriers and challenges in the short term but, after all, that is what engineers live for and there isn't anything that can't be resolved given the state of todays embedded systems and networking technologies.

However there are two key societal challenges – those of privacy/security and energy consumption. These are often lost in the current 'excitement' that surround IoT but ultimately these challenges will prove to be the key that determines the long-term sustainability of the Internet of Things. If you are looking for somewhere to make new contributions then these are worthwhile areas to consider and direct your efforts towards.

Acknowledgment

This research was funded under the Strategic Partnership Program of Science Foundation Ireland (SFI) and co-funded by SFI and FotoNation Ltd. Project ID: 13/SPP/I2868 on "*Next Generation Imaging for Smartphone and Embedded Platforms*".